\documentclass[preprint2]{aastex}

\shorttitle{Fossil Groups in the SDSS}
\shortauthors{Santos et al.}

\begin{document}


\title{Fossil Groups in the Sloan Digital Sky Survey}


\author{Walter A. Santos, Claudia Mendes de Oliveira, Laerte Sodr\'e Jr.}
\affil{Departamento de Astronomia, Instituto de Astronomia, Geof\'isica \\ e Ci\^encias Atmosf\'ericas da Universidade de S\~ao Paulo, \\
    Rua do Mat\~ao 1226, Cidade Universit\'aria, 05508-090, S\~ao Paulo, Brazil}
\email{walterjr@astro.iag.usp.br, oliveira@astro.iag.usp.br, laerte@astro.iag.usp.br}

\begin{abstract}

A search for fossil groups in the Sloan Digital Sky
Survey was performed using virtual observatory tools. 
A cross-match of the positions of all
SDSS Luminous Red Galaxies  (with r $<$ 19 and measured
spectroscopic redshifts) with sources in the ROSAT
All-Sky Survey catalog resulted in a list of 
elliptical  galaxies with extended X-ray emission 
(with a  galaxy/ROSAT-source distance
of less than 0.5 arcmin in all cases). A search for neighbors
of the selected elliptical galaxies  within a radius of $0.5 h_{70}^{-1}$ Mpc
was conducted taking into account the  r-band magnitudes and
spectroscopic or photometric redshifts of all objects within this
area, leading to a sample of 34 candidate fossil groups.
Considering this sample, the estimated space density of fossil systems is
$n =(1.0 \pm 0.6) \times 10^{-6}$ $h_{50}^3$ Mpc$^{-3}$.

\end{abstract}


\section{Introduction}

Fossil groups are systems with masses and X-ray luminosities comparable
to those of groups and clusters of galaxies, but whose light is dominated
by a single, isolated, large elliptical galaxy.  Their denomination,
``fossil'', comes from their possible formation scenario in which they may
have collapsed at an early epoch, being the oldest and most undisturbed
galaxy systems not yet absorbed by larger halos.

Studies of fossil groups started fairly recently, the first
system being identified by \citet{pon94}.
They defined a fossil group as a system with a bright 
($>$ 0.5 $\times$ 10$^{43}$ h$_{50}^{-2}$ erg\,s$^{-1}$)
and extended X-ray halo, dominated by one
brighter-than-L* elliptical galaxy which is surrounded by low-luminosity
companions, where the difference in magnitude between the bright dominant
elliptical and the next brightest companion is $>$ -2 mag.

Fossil groups may have been formed by the complete merging of galaxies 
within once normal groups/clusters \citep{vik99,mul99}. Indeed, this is
consistent with the fact that their baryon fraction is similar to that
observed in clusters \citep{mat05}.
It has been suggested that a fossil group could 
be a collapsed {\it compact group} of galaxies, but the connection between 
compact and fossil groups is not obvious.  From samples of today's compact 
groups, C. Mendes de Oliveira \& E. R. Carrasco (2007, in preparation) argue that compact groups which 
are in the process of merging are those with the poorest neighborhoods
and lowest velocity dispersions, not resembling the much more massive 
fossil group counterparts. X-ray studies of compact groups 
\citep{ebe94,pil95,pon96} have shown that although the
majority are X-ray loud (\citealt{pon96} infer that a 
fraction of 75\% of the sample of Hickson compact groups have hot intragroup gas),
their X-ray luminosities and temperatures are much lower than those 
measured for typical fossil groups. More recently, \citet{kho07}
have made a detailed comparison of the X-ray properties of fossil groups 
with those of other groups (see their Fig. 2) and have found that 
fossil groups have higher X-ray temperatures and luminosities for a given $L_R$  
when compared with normal groups.

Two studies
performed searches for fossil groups using well-defined selection criteria
\citep{vik99,jon03} and concluded that
these systems are quite abundant. There are, however, only
15 fossil groups known in the literature \citep[table 4 of][]{men06}.
The fact that so few such systems are identified to
date is not a surprise. They can easily be mistaken for isolated 
elliptical galaxies if spectroscopy of the member galaxies is not available, 
which is often the case.
There may also be some known clusters that can fall in the category of 
fossil groups, as discussed in Section 4.

The main contribution of this paper is to present a new
list of fossil groups, obtained from a search in the Sloan Digital Sky Survey database (SDSS Data Release 5 [DR5]),
which increases the number of such systems by a factor of 3 and
will allow statistical studies on fossil group properties to be done. The search
was performed using Structured Query Language and National Virtual
Observatory technologies (OpenSkyQuery and Astronomical Data Query Language).
We have made use of the ROSAT All-Sky Survey (RASS), since 
it is the only available X-ray sample with sensitivity and sky coverage large
enough to allow performing a search for fossil groups in the SDSS area
(although it is known to have limited sensitivity and variable flux limit
across the sky). 
Pointed observations with ROSAT or other satellites do not cover a large enough area. 

This paper is structured as follows. In Section 2 we present our
definition of a fossil system, as used for the search
performed in this paper.  Section 3
has our procedures and results, including a description of our selection
criteria and of the cross-match with the X-ray data. Section 4
presents a discussion of the results and perspectives. Finally, the
Appendix contains the details of the main queries of SDSS databases used
in this work.

\section{Our Definition of a Fossil System}

The definition of a fossil system adopted here is inspired by that
of \citet{jon03}, where the optical image consists of an elliptical
galaxy surrounded by fainter companions, so that the difference in the
R-band magnitude between the elliptical galaxy and the next brightest
companion  is $\ge$ -2 mag. The system
should also be detected as an extended X-ray source  (in the present
case, in the RASS catalog, described in \citealt{vog99}). 
However, unlike \citet{jon03} we do not impose a lower limit on X-ray luminosity.
We also consider SDSS r magnitudes, instead of the R-band magnitude adopted by those authors. 

Additionally, we consider
a fixed value for the search radius around the dominant elliptical
galaxy: $0.5 h_{70}^{-1}$ Mpc. \citet{jon03} consider a search radius
of half the projected virial radius ($0.5~r_{vir}$), which varies for each 
group. From \citet{kho07}, one can verify that, for the fossil groups studied so far, 
$0.5~r_{vir}$ varies between 0.22 and 0.68 Mpc, with a median 
value of 0.48 Mpc.

We do not assume any lower limit for the number of system members. Hence, even an
isolated elliptical galaxy may be classified as a fossil system, as long as it 
is associated with an extended X-ray source. 

\section{Procedure and Results}

In this section we describe all the steps we adopted to search for fossil systems
in the SDSS-DR5 \citep[see][]{ade07}.

The SDSS is a photometric and spectroscopic survey,
which provides data in a large area of the sky (mainly
in the north Galactic cap). It uses a dedicated 2.5 m telescope at
Apache Point Observatory in New Mexico. DR5
includes photometric data in five bands (ugriz) for 217 million objects
in an area over $8000~deg^{2}$, and 1,048,960 spectra of galaxies, quasars,
and stars selected from $5713~deg^{2}$ of the imaging data. DR5 contains
all the data from previous data releases and represents the conclusion
of the SDSS-I project. The SDSS spectroscopic data contain a magnitude-
and color-selected sample of luminous red galaxies (LRGs) for redshifts
up to 0.5, selected from cuts in color-magnitude space.

The search was performed in several steps, most of them using 
SQL (Structured Query Language) in the SDSS SQL database (CasJobs) and ADQL 
(Astronomical Data Query Language) in the NVO (National Virtual Observatory) 
tool OpenSkyQuery. The queries written in the languages SQL and ADQL are 
given in the Appendix.

\subsection{Elliptical galaxy selection} \label{ellipt_selec}

We selected galaxies from the LRG catalog \citep{eis01}. 
We consider only objects in this sample with $r < 19$. This restriction ensures that 
only galaxy companions with $r < 21$ are used to verify whether a system is a fossil
or not according to the definition above. The reason is that we are using 
photometric redshifts to identify the objects associated with the LRGs,
and, as discussed by \citet{csa03}, they are very uncertain for faint galaxies 
($r > 21$), making the identification of system members unreliable. 
These conditions yield 112,510 galaxies, which corresponds basically to the whole LRG 
catalog.

Note that not all objects in the LRG sample are ellipticals. Only at the end of
our procedure did we select systems dominated by elliptical galaxies through visual
inspection of the candidates.

\subsection{Cross-match with X-ray data} \label{xmatch}

The cross-match of our LRG sample with the RASS was performed with the function XMATCH in the NVO OpenSkyQuery tool\footnote{openskyquery.net/} 
\citep{omu05}, which 
allows us to cross-match astronomical catalogs using a general query language 
(ADQL, similar to SQL). One can also import a personal catalog of objects 
and cross-match it against other databases.

In order to obtain correct results from this cross-match, we had to take into 
account a limitation\footnote{http://openskyquery.net/Sky/skysite/help/limitations.aspx} of OpenSkyQuery:
the maximum number of objects for
cross-matches between query sets is currently 5000. Consequently, we 
have divided the 112,510 galaxy sample into subsets of up to 5000 objects. Then 
each of the subsets was imported into OpenSkyQuery, and the cross-matches 
were performed with one subset at a time.

The cross-match considers only objects that have extended X-ray emission 
as measured by RASS, i.e., when the source extent parameter \citep[described in][]{vog99} is larger than 0.
The function XMATCH has a parameter, set by the user, which specifies a 
confidence level for the positional coincidences of the objects in
both catalogs. For this parameter, we decided to use a value of 6$\sigma$
(because of the poor spatial resolution of the ROSAT data). In all cases we
checked the reliability of the cross-matched objects once we had the fossil
group candidates, and we noted that the distances between the LRG galaxies and
the ROSAT extended sources were always less than 0.5 arcmin (see Table 1).

The cross-match yielded 188 LRGs associated with extended 
X-ray sources. The next step, then, was to analyze the regions around these 
objects to determine whether they are indeed fossil systems.

\subsection{Selection of companions} \label{neighbors}

We have looked for LRG companions within radii of $0.5 h_{70}^{-1}$ Mpc 
(a flag in Table 1 indicates whether the structure would be classified as
a fossil if the search radius was $1 h_{70}^{-1}$ Mpc). 
The corresponding angular radii were computed
from the LRG spectroscopic redshifts (measured by SDSS) assuming a 
concordance, zero curvature cosmology with $h = 0.7$, $\Omega_{m} = 0.3$, 
and $\Omega_{\lambda} = 0.7$.

The companions of the cross-matched objects were found using a function in 
CasJobs called spGetNeighborsRadius, which has the following inputs for each
object in a list: identification, right ascension, declination and 
search-radius (in arcminutes). We have selected companions classified 
as galaxies by SDSS by considering only objects in the table "Galaxy," 
which is a subset of the SDSS database that has photometric data for 
galaxy-type objects.

We have used redshifts to identify the (putative) companions of our LRGs.
We used spectroscopic redshifts when available, but for the large majority
of the objects we had to use photometric redshifts. The latter,
as well as their uncertainties, are described in \citet{csa03}, 
and are now publicly available for all objects in SDSS-DR5 \citep{ade07}.

An object is identified as a companion of a LRG at redshift $z_{e}$ if its
redshift $z$ satisfies the condition:
\begin{equation}
z_e - \Delta z < z < z_{e} + \Delta z
\end{equation}
where $\Delta z$ is a range in redshift. This range could be, 
in principle, associated with the group velocity dispersion. However, 
since the majority of the neighbor galaxies only have photometric redshifts, 
for which the uncertainties are always larger than the expected velocity 
dispersions, we decided to adopt the photometric redshift
uncertainty (its median
value is 0.035) as the value for $\Delta z$.

   We adopted a value of
$\Delta z = 0.002$ for the few neighbors with spectroscopic redshifts. 
Note that this value is 
somewhat restrictive for the companions' selection, but since 
the final number of fossil system candidates in our sample is small (see
below), we decided to allow for false detections to avoid the exclusion 
of real fossil systems from our sample. In fact, we have repeated the search 
considering $\Delta z = 0.005$, which decreases the number of fossil systems
found at the end of the search (see discussion in section 4).

   We decided to consider only neighbors with photometric redshift
uncertainties smaller than or equal to 0.1, which is the mean 
error of photometric redshifts for objects with $r \sim 21$ \citep{csa03}, in
order to avoiding large values of these uncertainties affecting
our results. Here again we adopted an approach to avoid the exclusion 
of real fossil systems from the sample at the risk of including a few false ones.
The number of galaxies excluded from the analysis for the 34 groups selected, 
was 13\%, on average, of the total number of galaxies in these groups.

\subsection{Photometric condition for fossil groups} \label{photo_cond}

At this stage of the procedure we have a sample of LRGs with extended X-ray 
emission and surrounded by putative companions, constrained both spatially 
(in right ascension and declination) and in redshift. Now
we consider the photometric condition that defines fossil systems: 
the difference in magnitude in the r-band between an elliptical galaxy and the 
next brightest member should be at least 2 mag. For simplicity this condition
is verified for each selected neighbor
\begin{equation}
m_{i} > m_{1} + 2
\end{equation}
where $m_{1}$ and $m_{i}$ are the r-band magnitudes of the elliptical galaxy and 
of the i-th companion. If this condition is satisfied for every companion, the 
system is classified as a fossil group candidate by our search. From the 188 
LRGs, we found 44 that are in systems complying with this photometric condition.

\subsection{Analysis of the dominant galaxy} \label{verif}

Not all objects in the LRG sample are actually elliptical 
galaxies, so the next step was to do a visual inspection of the dominant galaxies
of these systems to verify whether they are indeed ellipticals. Among the 44 LRGs 
in the sample of fossil group candidates we found eight systems
which do not seem to have elliptical galaxy morphologies
(six are spirals and two are mergers), 
which were then excluded 
from our fossil group list.

One system had to be dropped from the list because its related X-ray emission turned out
to be associated with a strong X-ray white dwarf emitter, which appears as 
an extended source in
the ROSAT catalog.

One first-ranked elliptical galaxy in one of the systems was actually NGC 5846, which, 
according to the literature \citep{mah05}, is part of an isolated group. The
group, however, contains one elliptical 
galaxy (NGC 5813) that is more luminous, indicating that it cannot
be a fossil group. The search failed in this case due to the 
photometric redshift uncertainty of this member (greater than 0.1), which does not
have a spectroscopic redshift measured by the SDSS. 
This group was then also excluded from our list.

The remaining 34 fossil system candidates are presented in Table 1 and
Table 2. In Table 1, column 1 presents the fossil group ID number, column 2 the SDSS name of the dominant elliptical 
galaxy, column 3 the right ascension, column 4 the declination 
(the coordinates given in the latter two columns are for the central
elliptical galaxy, as measured by SDSS) and column 5 the ROSAT name for 
the source related to the system. 

In Table 2, column 1 lists the fossil group ID number,
column 2 the spectroscopic redshift of the elliptical 
galaxy, column 3 its r-band magnitude, column 4 the r-band absolute
magnitude, column 5 the radius in arcmin for 
$0.5 h_{70}^{-1}$ Mpc, column 6 the distance between the elliptical galaxy and the 
ROSAT source, column 7 the X-ray extent, in arcsec, as given in the ROSAT catalog,
column 8 gives the X-ray luminosity estimated from the ROSAT count rates, 
assuming a 
temperature $kT = 2$ keV and metallicity $Z = 0.4Z\sun$, column 9 lists relevant
references in the literature about the elliptical galaxy and column 10 lists relevant related
objects from other surveys or catalogs, from the NASA/IPAC Extragalactic Database (NED).

Note that, although it was not a requirement of the search process, 
all but one 
of the first-ranked elliptical galaxies (ID $=$ 28) 
have $L_{R} < L^{\star}_{R}$. 
This system has an r-band
absolute magnitude of $-21.25$ while the mean value of all the other systems is $-23.74$.

By comparing optical luminosities for the dominant elliptical galaxies of our fossil system
candidates with elliptical galaxies in low density environments \citep{col01,hel01,kel04} and in groups 
\citep{hic97,bal04,kel05,tan05,wei06,bld06},
we found that our dominant galaxies constitute the bright end of the elliptical galaxy magnitude distribution.
In particular, \citet{bal04} and \citet{kel05} show that there is a significant
population of luminous elliptical galaxies that are fairly isolated or in low density environments.
Therefore, \citet{bal04} conclude that some fraction of the red population must arise independently
of environment (e.g., by consumption of the internal gas supply) or they 
are fossil structures resulting from the complete
merging of bright galaxies in a group.

We have estimated the 
unabsorbed X-ray flux for each fossil system candidate from the count
rates in the ROSAT catalog using the tool 
WebPIMMS\footnote{http://heasarc.gsfc.nasa.gov/Tools/w3pimms.html,
WebPIMMS is a Web version of the PIMMS (v3.9b) tool. PIMMS was developed by Koji Mukai at the HEASARC. 
The first Web version was developed at the SAX Data Center. 
The SAX PIMMS package was ported to and modified for the HEASARC Web site by Michael Arida.} and considering
a Raymond-Smith model with a temperature of $kT = 2$ keV and metallicity 
$Z = 0.4~Z\sun$, which
represent average values for fossil groups \citep{kho07}. Note that our results
are not too sensitive to these assumptions. For example, if we use a temperature of $kT = 3$ keV instead, this changes the luminosities by less than 10\%. The dependence on the X-ray luminosity on the specific value of metallicity we choose is even weaker.

We have compared the estimated X-ray luminosities of our fossil system 
candidates with those of groups of galaxies  \citep{ebe94,pil95,hel00,mah00,
mul03}. We found that, although our
fossil system candidates show a large dispersion in X-ray luminosities, 
these tend to be considerably larger than the X-ray luminosities 
of groups of galaxies. In fact, the
X-ray luminosities of our fossil system candidates are, approximately, an order of magnitude larger,
on average, than what is found for both compact groups \citep{ebe94,pil95,hel00} and normal groups 
of galaxies \citep{hel00,mah00,mul03}.
On the other hand, our average value for the X-ray luminosity  ($53.8 \times 10^{42}  $ h$_{70}^{-2}$ erg s$^{-1}$) is similar to what
\citet{kho07} have found for 9 confirmed fossil groups 
($41.1 \times 10^{42} $ h$_{70}^{-2}$ erg s$^{-1}$, bolometric and measured within r$_{200}$). These results are in agreement 
with the fact that fossil groups have higher
X-ray luminosities, for a given optical luminosity, when compared with normal
groups of galaxies \citep[e.g.,][]{kho07}.

\subsection{Density of fossil groups} \label{density}

The density of fossil systems in the local universe may be computed with the 
$V_{max}$ technique introduced by \citet[see also \citealt{fel76}]{schm68}, which 
is appropriate for flux-limited samples. With this approach, the density 
is given by 
\begin{equation} 
n=\sum_i {1 \over V_{max,i}} 
\end{equation} 
and the associated statistical error is 
\begin{equation} 
\sigma_n= \left( \sum_i {1 \over V_{max,i}^2} \right)^{1/2} 
\end{equation} 
where $V_{max,i}$ represents the maximum comoving volume within which 
the fossil system would remain brighter than the sample limiting flux, 
and the sum is over all the systems in the sample. 

In the present case the detection of fossil systems is determined 
mostly by the X-rays observations, which are shallower than those in the 
optical.  Since RASS observations do not have a single limiting flux due 
to the different exposure times of observations across the sky, we have 
adopted the following procedure.  We first identified all extended sources 
in the RASS catalogue which were observed in fields with exposure times 
equal to that of each of the fossil groups listed in Table 1.  In order 
to find the flux limit for a given exposure time, we then estimated, for 
each source, the flux \citep[using flux2 - see][]{vog99} in the 0.1-2.4 
keV energy range, corrected for neutral hydrogen absorption. The flux 
limit corresponding to a given exposure time was then estimated as 
the peak of the histogram of fluxes of all sources found in images of the 
same exposure time. Correcting for the sky coverage of 
SDSS we obtain a space density of $n =(1.0 \pm 0.6) \times 10^{-6}$ 
$h_{50}^3$ Mpc$^{-3}$. The large error is due to the fact that  only 9 
systems in our sample have fluxes above the limit corresponding to their 
exposure times.

The density of fossil groups was also estimated by \citet{jon03}. 
From 5 systems with $L_X$(0.5-2 keV) $>~10^{42}~h_{50}^{-2}$ erg s$^{-1}$ they 
obtained a density $n \sim 4 \times 10^{-6}$ $h_{50}^3$ 
Mpc$^{-3}$, similar to ours. 

Another type of system related to fossil groups are the OLEGs- X-ray 
overluminous ($L_X$(0.5-2 keV) $>~2 \times 10^{43}~h_{50}^{-2}$ erg s$^{-1}$) 
elliptical galaxies \citep{vik99}. The number density of these 
systems estimated by \citet{vik99} is 
$n \simeq 2.4 \times 10^{-7}$ $h_{50}^3$ Mpc$^{-3}$.

\clearpage

\begin{deluxetable}{ccccc}
\tabletypesize{\scriptsize}
\tablecaption{Fossil Structures: Coordinates\label{tbl-1}}
\tablewidth{0pt}
\tablehead{
\colhead{ID} & \colhead{SDSS Name} & \colhead{SDSS RA} & \colhead{SDSS DEC} &
\colhead{ROSAT Name} \\
\colhead{Number} & \colhead{} &
\colhead{(J2000)} & \colhead{(J2000)} & \colhead{} 
}
\tablecolumns{5}
\startdata
1\tablenotemark{a} & J015021.27-100530.5 & 01:50:21.3 & -10:05:30.5 & J015021.2-100537 \\
2\tablenotemark{a}\tablenotemark{b} & J015241.95+010025.5 & 01:52:42.0 & +01:00:25.6 & J015242.1+010040 \\
3\tablenotemark{a} & J075244.19+455657.3 & 07:52:44.2 & +45:56:57.4 & J075243.6+455653 \\
4\tablenotemark{a} & J080730.75+340041.6 & 08:07:30.8 & +34:00:41.6 & J080730.4+340104 \\
5\tablenotemark{a}\tablenotemark{b} & J084257.55+362159.2 & 08:42:57.6 & +36:21:59.3 & J084257.7+362141 \\
6 & J084449.07+425642.1 & 08:44:56.6 & +42:58:35.7 & J084456.2+425826 \\
7\tablenotemark{a} & J090303.18+273929.3 & 09:03:03.2 & +27:39:29.4 & J090302.8+273939 \\
8\tablenotemark{a} & J094829.04+495506.7 & 09:48:29.0 & +49:55:06.7 & J094827.2+495523 \\
9 & J104302.57+005418.2 & 10:43:02.6 & +00:54:18.3 & J104303.0+005423 \\
10\tablenotemark{a}\tablenotemark{b} & J105452.03+552112.5 & 10:54:52.0 & +55:21:12.5 & J105453.3+552102 \\
11\tablenotemark{a}\tablenotemark{b} & J111439.76+403735.1 & 11:14:39.8 & +40:37:35.2 & J111439.4+403735 \\
12\tablenotemark{a} & J112155.27+104923.2 & 11:21:55.3 & +10:49:23.2 & J112154.2+104936 \\
13\tablenotemark{b} & J114128.29+055829.5 & 11:41:28.3 & +05:58:29.5 & J114128.4+055827 \\
14\tablenotemark{a}\tablenotemark{b}\tablenotemark{c} & J114647.57+095228.1 & 11:46:47.6 & +09:52:28.2 & J114647.4+095236 \\
15\tablenotemark{a} & J114803.81+565425.6 & 11:48:03.8 & +56:54:25.6 & J114804.1+565410 \\
16\tablenotemark{a} & J114915.02+481104.9 & 11:49:15.0 & +48:11:04.9 & J114912.9+481058 \\
17\tablenotemark{a}\tablenotemark{d} & J124742.07+413137.6 & 12:47:42.1 & +41:31:37.7 & J124740.1+413128 \\
18\tablenotemark{a}\tablenotemark{e} & J130009.36+444301.3 & 13:00:09.4 & +44:43:01.3 & J130007.7+444251 \\
19\tablenotemark{b} & J133559.98-033129.1 & 13:35:60.0 & -03:31:29.2 & J133601.0-033147 \\
20\tablenotemark{a} & J141004.19+414520.8 & 14:10:04.2 & +41:45:20.9 & J141006.4+414520 \\
21\tablenotemark{a} & J144516.86+003934.2 & 14:45:16.9 & +00:39:34.3 & J144516.1+003918 \\
22 & J145359.01+482417.1 & 14:53:59.0 & +48:24:17.1 & J145356.9+482418 \\
23 & J152946.28+440804.2 & 15:29:46.3 & +44:08:04.2 & J152947.4+440755 \\
24\tablenotemark{a} & J153344.13+033657.5 & 15:33:44.1 & +03:36:57.5 & J153344.0+033717 \\
25\tablenotemark{a} & J153950.78+304303.9 & 15:39:50.8 & +30:43:04.0 & J153950.3+304305 \\
26 & J154855.85+085044.3 & 15:48:55.9 & +08:50:44.4 & J154855.0+085102 \\
27\tablenotemark{b} & J161431.10+264350.3 & 16:14:31.1 & +26:43:50.4 & J161431.2+264336 \\
28\tablenotemark{a} & J163720.51+411120.2 & 16:37:20.5 & +41:11:20.3 & J163721.3+411106 \\
29\tablenotemark{a} & J164702.07+385004.2 & 16:47:02.1 & +38:50:04.3 & J164702.5+385003 \\
30\tablenotemark{a} & J171811.93+563956.1 & 17:18:11.9 & +56:39:56.1 & J171810.9+563955 \\
31\tablenotemark{a}\tablenotemark{b} & J172010.03+263732.0 & 17:20:10.0 & +26:37:32.1 & J172009.3+263727 \\
32\tablenotemark{a} & J172852.16+551640.8 & 17:28:52.2 & +55:16:40.8 & J172850.5+551622 \\
33\tablenotemark{a}\tablenotemark{f} & J225630.04-003210.8 & 22:56:30.0 & -00:32:10.7 & J225629.8-003231 \\
34\tablenotemark{a} & J235815.10+150543.5 & 23:58:15.1 & +15:05:43.6 & J235814.4+150524 \\
\enddata

\tablenotetext{a}{System is also classified as fossil considering $\Delta z = 0.005$, see text.}
\tablenotetext{b}{System is also classified as fossil considering radius = $1 h_{70}^{-1}$ Mpc, see text.}
\tablenotetext{c}{The elliptical galaxy for this system appears projected onto the 
cluster ZwCl 1144.3+1010 (it clearly does not belong to the cluster, according 
to its redshift). In this case, note that the X-Ray emission may be due to the cluster, 
instead of the fossil system candidate.}
\tablenotetext{d}{The elliptical galaxy is marked as blended,
but no deblending attemp was made by SDSS.}
\tablenotetext{e}{The elliptical galaxy is the result of a deblending procedure made in a
originally blended object, i. e., it had multiple peaks detected within it.}
\tablenotetext{f}{Same case as (c), the elliptical galaxy is projected onto the
cluster ABELL 2505.}

\end{deluxetable}

\clearpage

\begin{deluxetable}{cccccccccc}
\tabletypesize{\tiny}
\rotate
\tablecaption{Fossil Structures: Properties\label{tbl-2}}
\tablewidth{0pt}
\tablehead{
\colhead{ID} & \colhead{Redshift} & \colhead{r-band} & \colhead{r-band} &
\colhead{Radius\tablenotemark{a}} & \colhead{Distance\tablenotemark{b}} & \colhead{X-Ray ROSAT} & \colhead{L$_{X}$\tablenotemark{c} (h$_{70}^{-2}$)} &
\colhead{References} & \colhead{Related Objects} \\
\colhead{Number} & \colhead{} &
\colhead{Magnitude} & \colhead{Absolute Magnitude} & \colhead{(arcmin)} & \colhead{(arcmin)} 
& \colhead{Extent (arcsec)} & \colhead{(erg s$^{-1}$) ($0.5$-$2$keV)} & \colhead{} & \colhead{found in NED}
}
\tablecolumns{10}
\startdata

1 & 0.365 & 17.26 & -24.85 & 1.641 & 0.118 & 44 & 1.44E+44 & \nodata & \nodata \\
2 & 0.23 & 15.72 & -24.95 & 2.269 & 0.242 & 58 & 1.51E+44 & 1 & ABELL 0267 \\
3 & 0.052 & 14.46 & -22.64 & 8.241 & 0.123 & 10 & 5.62E+42 & 2 & 87GB\tablenotemark{d} 074906.3+460432 \\
4 & 0.208 & 16.38 & -24.05 & 2.449 & 0.388 & 16 & 4.21E+43 & 3 & FIRST\tablenotemark{e} J080730.7+340041 \\
5 & 0.282 & 16.79 & -24.52 & 1.951 & 0.306 & 63 & 2.95E+44 & 4 & ABELL 0697 \\
6 & 0.054 & 14.08 & -22.97 & 7.919 & 0.18 & 37 & 2.11E+42 & 5 & \nodata \\
7 & 0.489 & 18.06 & -25.19 & 1.38 & 0.182 & 37 & 1.74E+44 & \nodata & \nodata \\
8 & 0.409 & 18.21 & -24.22 & 1.529 & 0.407 & 10 & 6.26E+43 & \nodata & \nodata \\
9 & 0.125 & 15.98 & -23.16 & 3.704 & 0.138 & 25 & 4.99E+43 & 5 & \nodata \\
10 & 0.468 & 17.69 & -25.3 & 1.415 & 0.25 & 13 & 1.17E+44 & \nodata & 7C\tablenotemark{f} 1051+5537 \\
11 & 0.202 & 17.14 & -23.16 & 2.504 & 0.069 & 9 & 4.19E+43 & \nodata & \nodata \\
12 & 0.24 & 16.97 & -23.84 & 2.194 & 0.345 & 28 & 3.85E+43 & \nodata & \nodata \\
13 & 0.188 & 16.03 & -24.11 & 2.652 & 0.048 & 21 & 2.19E+43 & 3 & NVSS\tablenotemark{g} J114128+055828 \\
14 & 0.221 & 16.36 & -24.24 & 2.333 & 0.137 & 20 & 4.93E+43 & \nodata & \nodata \\
15 & 0.105 & 16.49 & -22.08 & 4.339 & 0.255 & 14 & 1.04E+43 & \nodata & \nodata \\
16 & 0.283 & 17.6 & -23.66 & 1.948 & 0.374 & 29 & 6.10E+43 & \nodata & \nodata \\
17 & 0.155 & 15.88 & -23.7 & 3.093 & 0.403 & 14 & 6.25E+42 & \nodata & \nodata \\
18 & 0.233 & 16.88 & -23.8 & 2.243 & 0.342 & 20 & 2.63E+43 & \nodata & \nodata \\
19 & 0.177 & 15.84 & -24.14 & 2.787 & 0.396 & 36 & 3.68E+43 & \nodata & \nodata \\
20 & 0.094 & 14.66 & -23.64 & 4.781 & 0.413 & 10 & 2.96E+42 & 5 & \nodata \\
21 & 0.306 & 17.94 & -23.65 & 1.843 & 0.326 & 35 & 7.21E+43 & \nodata & \nodata \\
22 & 0.146 & 15.87 & -23.58 & 3.255 & 0.351 & 29 & 5.75E+42 & \nodata & \nodata \\
23 & 0.148 & 15.77 & -23.71 & 3.224 & 0.252 & 13 & 1.10E+43 & 5 & \nodata \\
24 & 0.293 & 17.23 & -24.21 & 1.902 & 0.335 & 41 & 6.00E+43 & \nodata & \nodata \\
25 & 0.097 & 14.93 & -23.5 & 4.636 & 0.106 & 62 & 4.86E+43 & 6 & FIRST\tablenotemark{e} J153950.7+304303 , ABELL 2110 \\
26 & 0.072 & 13.5 & -24.25 & 6.06 & 0.362 & 59 & 5.09E+42 & 7 & UZC\tablenotemark{h} J154855.9+085045 \\
27 & 0.184 & 15.76 & -24.36 & 2.693 & 0.232 & 19 & 2.37E+43 & \nodata & \nodata \\
28 & 0.032 & 14.51 & -21.25 & 13.153 & 0.274 & 17 & 4.16E+41 & 8 & \nodata \\
29 & 0.135 & 16.13 & -23.09 & 3.48 & 0.083 & 27 & 6.87E+42 & 6 & FIRST\tablenotemark{e} J164702.0+385005 \\
30 & 0.114 & 15.27 & -23.55 & 4.037 & 0.144 & 68 & 4.67E+43 & 5, 3, 9, 10 & ZwCl\tablenotemark{i} 1717.9+5636 , 7C\tablenotemark{f} 1717+5643 \\
31 & 0.159 & 15.44 & -24.33 & 3.03 & 0.181 & 62 & 1.77E+44 & 5, 3, 10 & \nodata \\
32 & 0.148 & 16.72 & -22.81 & 3.214 & 0.393 & 14 & 3.53E+42 & \nodata & \nodata \\
33 & 0.224 & 16.81 & -23.91 & 2.314 & 0.351 & 15 & 2.18E+43 & 6 & FIRST\tablenotemark{e} J225630.0-003209 \\
34 & 0.178 & 16.08 & -23.98 & 2.763 & 0.369 & 21 & 9.26E+42 & \nodata & \nodata \\
\enddata

\tablenotetext{a}{Equivalent to the radius $0.5 h_{70}^{-1}$ Mpc.}
\tablenotetext{b}{Distance between the SDSS elliptical galaxy and its ROSAT counterpart.}
\tablenotetext{c}{X-ray luminosity estimated from ROSAT count rates, assuming a temperature $kT = 2$ keV and metallicity $Z = 0.4Z\sun$.}
\tablenotetext{d}{1987 Green Bank Radio Survey.}
\tablenotetext{e}{Faint Images of the Radio Sky at Twenty Centimeters.}
\tablenotetext{f}{Seventh Cambridge Radio Catalog.}
\tablenotetext{g}{NRAO VLA Sky Survey.}
\tablenotetext{h}{Updated Zwicky Catalog.}
\tablenotetext{i}{Zwicky's Catalog of Galaxies and Clusters of Galaxies.}

\tablerefs{
(1) \citealt{pli05}; (2) \citealt{ver01}; (3) \citealt{bes05};
(4) \citealt{met00}; (5) \citealt{mer05}; (6) \citealt{bri00};
(7) \citealt{fal99}; (8) \citealt{rin02}; (9) \citealt{hao05};
(10) \citealt{pra03}.}

\end{deluxetable}

\clearpage

\section{Discussion}

The search for fossil groups performed in this work was intended to be as inclusive as possible, 
minimizing the probability of exclusion of real fossil systems
from our sample. It should then be noted that 
our procedure leads to an increase in the
probability of having false fossil systems in our final list. 
The conditions considered throughout the search, such as small values for
the system radius and for the minimum redshift range constraint,
tend to decrease the number of companions 
around the elliptical galaxies, 
which in turn increases the probability of fossil system detection, 
since there are fewer galaxies subjected to the photometric 
condition for a system to be a fossil. For the final results, 
we decided to adopt $0.5 h_{70}^{-1}$ Mpc for the radius around the luminous 
elliptical galaxies and 0.002 for the minimum redshift range constraint. 
This gave us a final list with 34 groups, which are listed
in tables 1 and 2.

We have re-done the same procedure considering another value for the 
search radius, namely $1 h_{70}^{-1}$ Mpc. 
We found that the search yields only 9 
fossil systems for $1 h_{70}^{-1}$ Mpc and when the redshift range is
0.002. A total of 26 fossil systems 
are found when we consider $0.5 h_{70}^{-1}$ Mpc and, instead, a minimum 
redshift range of 0.005. For the most restrictive case, i. e., a search radius of $1 h_{70}^{-1}$ Mpc 
and the minimum redshift range of 0.005, we only find 6 fossil systems.
Therefore, the number of fossil system candidates depends critically on these 
parameters.  Table 1, which presents the fossil system candidates, shows flags indicating 
whether the system is also classified as a fossil in those two other scenarios.

In this work, in order to identify galaxies spatially close to the  elliptical galaxy, we used 
photometric redshifts, for which uncertainties are known to be high in comparison with those of 
spectroscopic redshifts. Thus, even considering only companion galaxies with r-band 
magnitude less than 21 and excluding galaxies with photometric redshift uncertainty larger than 
0.1 (see section 3.3), 
the photometric redshift accuracy does limit the efficiency of 
the search. However, for the kind of search conducted here, 
where a large survey such as SDSS is required, the use of photometric redshifts is the only viable 
way to find structures around the elliptical galaxies.

We would like to note that no redshift cut was applied
to the sample. The highest redshift fossil group found was at z=0.47
(system with ID $=$ 10).

\subsection{Other classifications for the fossil groups} \label{ned_search}

We performed a search in the
NED\footnote{http://nedwww.ipac.caltech.edu/} 
for each of the 34 fossil system candidates found in this work, in order to
find additional information on them. Specifically, we looked for related objects
in other surveys/catalogs and scientific papers that cite the elliptical galaxies of the fossil
systems. These are listed in Table 2.

We found that eight fossil system candidates (IDs = 3, 4, 10, 13, 25, 29, 30, 33)
are related to radio emission sources, from different surveys (listed in Table 2). Four
systems (IDs = 4, 13, 30, 31) were classified as radio-loud active galactic nuclei (AGNs), described in
\citet{bes05}. One of the systems (ID = 30) was also selected as an AGN in \citet{hao05}. In addition, one system (ID = 3) is part
of a catalogue of quasars and active nuclei, described in \citet{ver01}. It is worth noting that
the soft X-ray luminosity of these fossil system candidates might be contaminated by
non resolved central AGNs.

Six systems (IDs = 6, 9, 20, 23, 30, 31), out of the 34 fossil system
candidates, were classified as
galaxy groups by \citet{mer05}. In that paper, they used a spectroscopic galaxy
sample from SDSS-DR3 (Third Data Release) to produce a group catalog, using a method
based on the friends-of-friends algorithm.

We also found that four systems from our list (IDs = 2, 5, 26, 30) may belong to 
galaxy clusters. In particular, our search in NED indicates that system 
5 is related
to the cluster ABELL 0697 and system 2 
is related to ABELL 0267. The systems with
IDs = 26, 30 are classified as clusters in the Zwicky catalog \citep{fal99}.

\subsection{Cross-check with the list of previously known fossil groups} \label{cross_check}

We have cross-checked our results with the list of 
known fossil groups presented in \citet{men06}. 
Out of the 15 known fossil groups, SDSS has no available data for six of them,
and in seven others their main elliptical galaxies had no
measured spectra in SDSS; thus, they were not selected in the first step of the
search. The elliptical galaxies of the remaining two (RX J1159.8+5531;
RX J1340.6+4018) were classified as LRG with measured
spectroscopic redshifts, but were not selected in the cross-match with
ROSAT performed in OpenSkyQuery.

We decided to search in the RASS catalog for X-ray sources  
related to those two known fossil groups, but in an independent way. We used
ROSAT tool Source Browser\footnote{http://www.xray.mpe.mpg.de/cgi-bin/rosat/src-browser} for the search, using the coordinates given in \citet{men06}.

For the fossil group RX J1159.8+5531 \citep{vik99}, at $11^{h}59^{m}51\fs 4, 55\degr 32\arcmin 01 \arcsec$,
the nearest extended source has the coordinates $11^{h}57^{m}56\fs 10, 55\degr 27\arcmin 17\farcs 5$, which
gives a distance of $\sim 17$ arcmin, much greater than expected for the occurence of a cross-match in OpenSkyQuery. However,
there is a nearer (distance $\sim 0.6$ arcmin) point-like source (ext = 0) at $11^{h}59^{m}55\fs 40, 55\degr 31\arcmin 53\farcs 0$.
In the original paper \citep{vik99}, they found that the isolated elliptical
galaxy lies exactly at the peak of the X-ray emission (see their Fig. 1).

For RX J1340.6+4018 \citep{pon94}, whose coordinates are
$13^{h}40^{m}33\fs 4, 40\degr 17\arcmin 48 \arcsec$, we found the nearest extended source
at $13^{h}40^{m}38\fs 10, 40\degr 36\arcmin 39\farcs 5$, which represents a distance of $\sim 19$ arcmin, again
too large to have a cross-match in OpenSkyQuery. The nearest point-like source is at a distance of $\sim 7.6$ arcmin, at
$13^{h}41^{m}02\fs 60, 40\degr 12\arcmin 38\farcs 0$. According to \citet{pon94}, the X-ray emission
centroid coincides with the elliptical galaxy within 10 arcseconds (see their Fig. 1).

The discrepancies found between RASS extended X-ray sources and the X-ray emission
detected for those known fossil groups may be due to differences in spatial resolution
between the ROSAT All-Sky Survey catalog and the pointed observations
performed by those authors \citep{pon94,vik99}. The uncertainties of
the extended source centers in the ROSAT catalog can also affect our
cross-match results.

In addition, even if the cross-match had been succesful for RX J1340.6+4018, it still would not be considered a fossil system
in our search. According to SDSS, there is a
bright galaxy with measured concordant spectrocopic redshift at 
$13^{h}40^{m}37\fs 64, 40\degr 15\arcmin 16\farcs 3$, thus inside the radius of
$0.5 h_{70}^{-1}$ Mpc (distant $\sim 2.6$ arcmin). 
The difference in r-band between this galaxy and the central one is 1.3 and $\Delta z = 0.002$.
However, when considering the original definition of fossil groups by \citet{pon94} and 
\citet{jon03}, RX J1340.6+4018 is indeed a fossil system
because that galaxy (at $13^{h}40^{m}37\fs 64, 40\degr 15\arcmin 16\farcs 3$) lies outside
the half virial radius (around 270 kpc in this case).

\subsection{Perspectives} \label{perspectives}

There are several open questions in the study of the formation and
evolution of fossil groups, which can be tackled using the sample
cataloged here. We discuss below two studies which can be
performed using our sample:
(1) the determination of
ages, metallicities and abundances of the first-ranked galaxies in 
fossil groups and (2) the determination of the  luminosity function
of fossil group galaxies.
Motivation and more details of these studies
are given in the following.

\begin{itemize}

\item (1) The morphology of the first-ranked galaxies in seven fossil groups was
investigated by \citet{kho06a},
from R-band and near infrared images and none
of the galaxies were found to have boxy isophotes, in contrast with
results found for brightest cluster galaxies (BCGs). This suggests
that the brightest galaxies in fossil groups may have a different structure
than that of BCGs. In fact, \citet[astro-ph/0702095]{kho07} concluded
that there is an absence of recent merging in fossil groups and
\citet{don05} suggested that fossil groups have assembled 50\% of their masses at z
$>$ 1 and have grown through minor mergers at later stages. It would be
most useful to have determinations of ages/metallicities and alpha
enhancements of the central elliptical galaxies of fossil  groups but
so far no spectroscopic analysis of such galaxies has been made.
Our sample offers a prime chance to do such an analysis, since spectra
are available for all first-ranked galaxies of the 34 fossil groups
catalogued here through the SDSS database. 

\item (2)  \citet{don04}
argued that fossil groups have two orders of magnitude less substructure
than predicted in CDM cosmological simulations. The significance of
this result, however, depends on a reliable determination of the galaxy
luminosity function of fossil groups, which was not available at the
time of that study.  Recently, two rich fossil groups,
RX~J1552.2+2013 and RX~J1416.4+2315$\,$
have been studied spectroscopically and their luminosity functions
determined down to M=-18 \citep{men06,cyp06}.  For RX~J1552.2+2013$\,$
the luminosity function shows a
lack of faint galaxies, with $\alpha = -0.6$.  For RX~J1416.4+2315$\,$
the spectroscopic
luminosity function was measured with lower accuracy, to have a value with
the faint end well fit by $\alpha = -1.2$, compared with $\alpha = -0.6$ measured
for the LF of the same group \citep[by][]{kho06b}. More fossil groups
have to be studied in detail for a better understanding of the shapes of
the luminosity functions of these systems, specially at the faint end.
The sample cataloged here may be useful for such studies, as 
more measurements of redshifts for the possible group members are obtained.

\end{itemize}

\acknowledgments

The authors would like to acknowledge support from the Brazilian agencies
FAPESP (thematic project 01/07342-7, MSc scholarship 05/01019-0) and CNPq. We
would like to thank Gast\~ao Lima Neto, Trevor Ponman, Eduardo Cypriano, and Abilio Mateus Jr.
for useful discussions, as well as the referee for his/her comments, that helped
us to improve the paper. WAS would like to thank 
the organization of the NVO School, which took place in Aspen, USA, in September/2006, 
for finantial support for the conference and providing 
motivation and tools to perform this work, as well as the following
participants of the NVO project team at the school: Omar Lopez-Cruz, Takayuki Tamura, and
Don Lindler. WAS would also like to thank the SDSS/VO team at Johns Hopkins University, 
for finantial support and hospitality during exchange programs and
in particular Alex Szalay, Ani Thakar, Tamas Budavari, and Maria Nieto-Santisteban for useful discussions.
LJS has benefited from CAPES/Cofecub financial support and would like
to thank the hospitality  of the Institut d'Astrophysique
de Paris, where part of this work was developed.

This research has made use of software provided by the US National Virtual Observatory, which is sponsored by the National Science Foundation.

Funding for the SDSS and SDSS-II has been provided by the Alfred P. Sloan Foundation, the Participating Institutions, the National Science Foundation, the U.S. Department of Energy, the National Aeronautics and Space Administration, the Japanese Monbukagakusho, the Max Planck Society, and the Higher Education Funding Council for England. The SDSS Web Site is http://www.sdss.org/.

The SDSS is managed by the Astrophysical Research Consortium for the Participating Institutions. The Participating Institutions are the American Museum of Natural History, Astrophysical Institute Potsdam, University of Basel, University of Cambridge, Case Western Reserve University, University of Chicago, Drexel University, Fermilab, the Institute for Advanced Study, the Japan Participation Group, Johns Hopkins University, the Joint Institute for Nuclear Astrophysics, the Kavli Institute for Particle Astrophysics and Cosmology, the Korean Scientist Group, the Chinese Academy of Sciences (LAMOST), Los Alamos National Laboratory, the Max-Planck-Institute for Astronomy (MPIA), the Max-Planck-Institute for Astrophysics (MPA), New Mexico State University, Ohio State University, University of Pittsburgh, University of Portsmouth, Princeton University, the United States Naval Observatory, and the University of Washington.

\appendix

\section{SQL and ADQL Queries}

Note that only the main queries are listed here.

\vspace{0.5cm}

\noindent {\bf SQL query to select luminous red galaxies from SDSS:}

{\tt
\noindent SELECT \\
\hspace*{0.5cm} p.objID, s.ra, s.dec, s.z as redshift, p.u, p.g, p.r, p.i, p.z, \\ 
\hspace*{0.5cm} (p.u-p.r) as u\_r, l.ew as d4000, s.eClass, p.lnlDev\_r, \\
\hspace*{0.5cm} p.lnlExp\_r INTO lrgs \\
FROM SpecObj as s, Galaxy as p, SpecLineIndex as l \\
WHERE \\
\hspace*{0.5cm} s.bestObjID = p.objID AND \\
\hspace*{0.5cm} s.specClass = 2 AND \\
\hspace*{0.5cm} s.zStatus $>$ 1 AND \\
\hspace*{0.5cm} (s.primTarget \& 0x00000020 $>$ 0) AND \\
\hspace*{0.5cm} s.z $>$ 0 AND p.r $<$ 19 AND \\
\hspace*{0.5cm} l.specobjid = s.specobjid AND \\
\hspace*{0.5cm} l.name = `4000Abreak' \\
ORDER BY p.objID
}

\vspace{0.5cm}

\noindent {\bf ADQL query on OpenSkyQuery that performs the cross-match between
LRG galaxies and ROSAT extended objetcs:}

{\tt
\noindent SELECT x.objid, x.ra, x.dec, x.ext, t.* \\
FROM Rosat:PhotoPrimary x, MyData:lrgs\_1 t \\
WHERE XMATCH(x, t) $<$ 6 AND x.ext $>$ 0
}

\vspace{0.5cm}

\noindent {\bf SQL queries to select neighbors around the elliptical galaxies:}

{\tt
\noindent CREATE TABLE neighbors ( \\
\hspace*{0.5cm} ra float, \\
\hspace*{0.5cm} dec float, \\
\hspace*{0.5cm} rad float, \\
\hspace*{0.5cm} id int, \\
\hspace*{0.5cm} z float, \\
\hspace*{0.5cm} objid bigint, \\
\hspace*{0.5cm} r real \\
)

\noindent CREATE TABLE \#UPLOAD( \\
\hspace*{0.5cm} up\_ra FLOAT, \\
\hspace*{0.5cm} up\_dec FLOAT, \\
\hspace*{0.5cm} up\_rad FLOAT, \\
\hspace*{0.5cm} up\_id int \\
) \\
INSERT INTO \#UPLOAD \\
SELECT \\
\hspace*{0.5cm} ra AS UP\_RA, \\
\hspace*{0.5cm} dec AS UP\_DEC, \\
\hspace*{0.5cm} rad as UP\_RAD, \\
\hspace*{0.5cm} id AS UP\_ID \\
FROM mydb.radius \\
ORDER BY id

\noindent CREATE TABLE \#tmp ( \\
\hspace*{0.5cm} up\_id int, \\
\hspace*{0.5cm} objid bigint \\
) \\
INSERT INTO \#tmp \\
EXEC spGetNeighborsRadius \\

\noindent INSERT INTO mydb.neighbors \\
SELECT a.ra, a.dec, a.rad, a.id, a.redshift, t.objid, g.r \\
FROM \#tmp t, mydb.radius a, Galaxy g \\
WHERE t.up\_id = a.id AND t.objid = g.objID AND g.r $<$ 21 \\
ORDER BY a.id, t.objid
}

\vspace{0.5cm}

\noindent {\bf SQL Query to select photometric redshifts of the neighbors:}

{\tt
\noindent SELECT n.id, n.objid, p.z, p.zErr, p.quality, n.r \\
INTO redshifts \\
FROM neighbors as n, dr5.photoz as p \\
WHERE n.objid = p.objid \\
ORDER BY n.id, n.objid
}

\vspace{0.5cm}

\noindent {\bf SQL queries to replace photometric redshifts with spectroscopic ones when available:}

{\tt
\noindent UPDATE redshifts \\
SET z = \\
\hspace*{0.5cm} (SELECT TOP 1 d.z \\
\hspace*{0.5cm}	FROM dr5.specObj as d \\
\hspace*{0.5cm}	WHERE d.bestObjID = redshifts.objID AND d.zStatus $>$ 1 AND d.z $>$ 0) \\
WHERE EXISTS \\
\hspace*{0.5cm} (SELECT TOP 1 d.z \\
\hspace*{0.5cm}	FROM dr5.specObj as d \\
\hspace*{0.5cm}	WHERE d.bestObjID = redshifts.objID AND d.zStatus $>$ 1 AND d.z $>$ 0)

\noindent UPDATE redshifts \\
SET zErr = 0.002 \\
WHERE EXISTS \\
\hspace*{0.5cm} (SELECT top 1 d.zErr \\
\hspace*{0.5cm}	FROM dr5.specObj as d \\
\hspace*{0.5cm}	WHERE d.bestObjID = redshifts.objID AND d.zStatus $>$ 1 AND d.z $>$ 0)
}

\vspace{0.5cm}

\noindent {\bf SQL query to exclude objects that have large redshift uncertainty:}

{\tt
\noindent DELETE \\
FROM redshifts \\
WHERE zErr $>$ 0.1
}

\vspace{0.5cm}

\noindent {\bf SQL query to constrain systems using redshifts:}

{\tt
\noindent SELECT s.* INTO groups FROM redshifts as s, lrgs as t \\
WHERE \\
\hspace*{0.5cm} s.id = t.id AND \\
\hspace*{0.5cm} (s.z BETWEEN (t.redshift - s.zErr) AND (t.redshift + s.zErr)) \\
ORDER BY s.id, s.objid
}

\vspace{0.5cm}

\noindent {\bf SQL query to identify non-fossil systems:}

{\tt
\noindent SELECT s.* INTO nofossils \\
FROM lrgs as l, groups as s \\
WHERE s.id = l.id AND (s.r $<=$ (l.r + 2)) \\
ORDER BY s.id, s.objid
}

\vspace{0.5cm}

\noindent {\bf SQL query that deletes non-fossil systems, leaving only
fossil systems candidates in the table:}

{\tt
\noindent DELETE FROM fossils \\
WHERE EXISTS \\
\hspace*{0.5cm} (SELECT * \\
\hspace*{0.5cm} FROM nofossils as g \\
\hspace*{0.5cm} WHERE g.id = fossils.id)
}


\begin{thebibliography}{}
\bibitem[Adelman-McCarthy et al.(2007)]{ade07} Adelman-McCarthy, J. K., et al.  2007, \apjs, submitted
\bibitem[Baldry et al.(2006)]{bld06} Baldry, I. K., Balogh, M. L., Bower, R. G., Glazebrook, K., 
Nichol, R. C., Bamford, S. P., \& Budavari, T.  2006, \mnras, 373, 469
\bibitem[Balogh et al.(2004)]{bal04} Balogh, M. L., Baldry, I. K., Nichol, R., Miller, C., Bower, R., \& Glazebrook, K.  2004, \apj, 615, 101
\bibitem[Best et al.(2005)]{bes05} Best, P. N., Kauffmann, G., Heckman, T. M., \& Ivezic, Z.  2005, \mnras, 362, 9
\bibitem[Blanton \& Roweis(2007)]{bla07} Blanton, M. R., \& Roweis, S.  2007, \aj, 133, 734
\bibitem[Brinkmann et al.(2000)]{bri00} Brinkmann, W., Laurent-Muehleisen, S. A., Voges, W., Siebert, J., 
Becker, R. H., Brotherton, M. S., White, R. L., \& Gregg, M. D.  2000, \aap, 356, 445
\bibitem[Colbert et al.(2001)]{col01} Colbert, J. W., Mulchaey, J. S., \& Zabludoff, A. I.  2001, \aj, 121, 808
\bibitem[Csabai et al.(2003)]{csa03} Csabai, I., et al.  2003, \aj, 125, 580
\bibitem[Cypriano et al.(2006)]{cyp06} Cypriano, E. S., Mendes de Oliveira, C. L., Sodré, L., Jr.  2006, \aj, 132, 514
\bibitem[D'Onghia \& Lake(2004)]{don04} D'Onghia, E., \& Lake, G.  2004, \apj, 612, 628
\bibitem[D'Onghia et al.(2005)]{don05} D'Onghia, E., Sommer-Larsen, J., Romeo, A. D., Burkert, A., 
Pedersen, K., Portinari, L., \& Rasmussen, J.  2005, \apj, 630, 109
\bibitem[Ebeling et al.(1994)]{ebe94} Ebeling, H., Voges, W., \& Boehringer, H.  1994, \apj, 436, 44
\bibitem[Eisenstein et al.(2001)]{eis01} Eisenstein, D. J., et al.  2001, \aj, 122, 2267
\bibitem[Falco et al.(1999)]{fal99} Falco, E. E., et al.  1999, \pasp, 111, 438
\bibitem[Felten(1976)]{fel76} Felten, J. E.  1976, \apj, 207, 700
\bibitem[Hao et al.(2005)]{hao05} Hao, L., et al.  2005, \aj, 129, 1783
\bibitem[Helsdon \& Ponman(2000)]{hel00} Helsdon, S. F., \& Ponman, T, J.  2000, \mnras, 319, 933
\bibitem[Helsdon et al.(2001)]{hel01} Helsdon, S. F., Ponman, T. J., O'Sullivan, E., \& Forbes, D. A.  2001, \mnras, 325, 693
\bibitem[Hickson(1997)]{hic97} Hickson, P.  1997, \araa, 35, 357
\bibitem[Jones et al.(2003)]{jon03} Jones, L. R., Ponman, T. J., Horton, A., Babul,
A., Ebeling, H., \& Burke, D. J.  2003, \mnras, 343, 627
\bibitem[Kelm \& Focardi(2004)]{kel04} Kelm, B., \& Focardi, P.  2004, \aap, 418, 937
\bibitem[Kelm et al.(2005)]{kel05} Kelm, B., Focardi, P., \& Sorrentino, G.  2005, \aap, 442, 117
\bibitem[Khosroshahi et al.(2006a)]{kho06a} Khosroshahi, H. G., Ponman, T. J., \& Jones, L. R.  2006, \mnras, 372, L68
\bibitem[Khosroshahi et al.(2006b)]{kho06b} Khosroshahi, H. G., Maughan, B. J., 
Ponman, T. J., \& Jones, L. R.  2006, \mnras, 369, 1211
\bibitem[Khosroshahi et al.(2007)]{kho07} Khosroshahi, H. G., Ponman, T. J., \& Jones, L. R.  2007, \mnras, in press (astro-ph/0702095)
\bibitem[Mahdavi et al.(2000)]{mah00} Mahdavi, A., Böhringer, H., Geller, M. J., \& Ramella, M.  2000, \apj, 534, 114
\bibitem[Mahdavi et al.(2005)]{mah05} Mahdavi, A., Trentham, N., \& Tully, R. B.  2005, \aj, 130, 1502
\bibitem[Mathews et al.(2005)]{mat05} Mathews, W. G., Faltenbacher, A., Brighenti, F., \& Buote, D. A.  2005, \apj, 634, 137
\bibitem[Mendes de Oliveira et al.(2006)]{men06} Mendes de Oliveira, C., Cypriano, E. S., \& Sodr\'e, L., Jr.  2006, \aj, 131, 158
\bibitem[Merch\'an \& Zandivarez(2005)]{mer05} Merch\'an, M. E., \& Zandivarez, A.  2005, \apj, 630, 759
\bibitem[Metzger \& Ma(2000)]{met00} Metzger, M. R., \& Ma, C.-P.  2000, \aj, 120, 2879
\bibitem[Mulchaey \& Zabludoff(1999)]{mul99} Mulchaey, J. S., \& Zabludoff, A. I.  1999, \apj, 514, 133
\bibitem[Mulchaey et al.(2003)]{mul03} Mulchaey, J. S., Davis, D. S., Mushotzky, R. F., \& Burstein, D.  2003, \apjs, 145, 39
\bibitem[O'Mullane et al.(2003)]{omu05} O'Mullane, W., et al.  2005, in ASP Conf. Ser. 347,
Astronomical Data Analysis Software and Systems XIV, ed. P. Shopbell, M. Britton, \& R. Ebert
(San Francisco: ASP), 341
\bibitem[Pildis et al.(1995)]{pil95} Pildis, R. A., Bregman, J. N., \& Evrard, A. E.  1995, \apj, 443, 514
\bibitem[Plionis et al.(2005)]{pli05} Plionis, M., Basilakos, S., Georgantopoulos, I., \& Georgakakis, A.  2005, \apj, 622, 17
\bibitem[Ponman et al.(1994)]{pon94} Ponman, T. J., Allan, D. J., Jones, L. R., Merrifield, M., 
McHardy, I. M., Lehto, H. J., \& Luppino, G. A.  1994, \nat, 369, 462
\bibitem[Ponman et al.(1996)]{pon96} Ponman, T. J., Bourner, P. D. J., Ebeling, H., \& Bohringer, H.  1996, \mnras, 283, 690
\bibitem[Prada et al.(2003)]{pra03} Prada, F., et al.  2003, \apj, 598, 260
\bibitem[Rines et al.(2002)]{rin02} Rines, K., Geller, M. J., Diaferio, A., Mahdavi, A., Mohr, J. J., \& 
Wegner, G.  2002, \aj, 124, 1266
\bibitem[Schlegel et al.(1998)]{sch98} Schlegel, D. J., Finkbeiner, D. P., \& Davis, M.  1998, \apj, 500, 525
\bibitem[Schmidt(1968)]{schm68} Schmidt, M.  1968, \apj, 151, 393
\bibitem[Tanaka et al.(2005)]{tan05} Tanaka, M., Kodama, T., Arimoto, N., Okamura, S., Umetsu, K., 
Shimasaku, K., Tanaka, I., \& Yamada, T.  2005, \mnras, 362, 268
\bibitem[V\'eron-Cetty \& V\'eron(2001)]{ver01} V\'eron-Cetty, M.-P., \& V\'eron, P.  2001, \aap, 374, 92
\bibitem[Vikhlinin et al.(1999)]{vik99} Vikhlinin, A., McNamara, B. R., Hornstrup, A., Quintana, H., 
Forman, W., Jones, C., \& Way, M.  1999, \apj, 520, L1
\bibitem[Voges et al.(1999)]{vog99} Voges, W., et al.  1999, \aap, 349, 389
\bibitem[Weinmann et al.(2006)]{wei06} Weinmann, S. M., van den Bosch, F. C., Yang, X., \& Mo, H. J.  2006, \mnras, 366, 2

\end{thebibliography}
\end{document}